# First-principles investigation of interfacial reconstruction in epitaxial SrTiO$_3$/Si photocathodes

Wen-Yi Tong[1], Eric Bousquet[1], Matjaž Spreitzer[2] and Philippe Ghosez[1,*]

[1] *Theoretical Materials Physics, Q-MAT, CESAM, Université de Liège, Sart Tilman B-4000, Belgium*
[2] *Advanced Materials Department, Jozef Stefan Institute, Jamova 39, 1000 Ljubljana, Slovenia*

Epitaxial SrTiO$_3$ (STO) on Si is nowadays the benchmark initial platform for the further addition of functional oxides on Si. Starting the growth of STO on a Sr-passivated Si substrate with ½ monolayer (ML) Sr coverage and a (1 × 2) reconstructed Si surface with rows of Si dimers, the final STO/Sr/Si stack exhibits 1 ML Sr coverage and a (1 × 1) Si surface without dimer. Using first-principles density functional theory calculations, we investigate how the interface evolves from ½ ML to 1 ML Sr coverage, concluding that the latter is indeed most stable and that the reconstruction of the interface takes place during the early stage of the layer-by-layer deposition. Going further, we determine the band alignment of the final stable interface and assess its potential interest as photocathode for water reduction.

Since the pioneering work of McKee in 1998 [1], the epitaxial growth of perovskite oxide films on silicon has generated continuous interest [2-4] due to the perspectives it offers to integrate the various functionalities of perovskite oxides [5,6] directly into electronic devices [7]. In this context, many efforts [8,9] have been devoted to the integration of SrTiO$_3$ (STO) on Si. STO, being one of the most popular substrates for the growth of perovskite films, appears as the ideal prototypical template for the further deposition of functional perovskites on Si [10-13]. Moreover, Ji *et al.* [14] showed that epitaxial integration of STO on Si provides stable photocurrents, demonstrating also the potential interest for electrochemical water splitting.

To date, high-quality STO/Si heterostructures can be successfully grown by molecular beam epitaxy [1,15,16], atomic layer deposition [17], pulsed laser deposition [18,19], or other layer-by-layer growth techniques [20]. However, direct growth of STO films on a Si (001) substrate is impossible since STO will react with Si during the initial stage and form an amorphous transition layer at the interface that affects the epitaxy. An appropriate buffer layer is therefore necessary to prevent Si from being oxidized. Till now, numerous atomic structures have been proposed as buffer layers both theoretically [21-28] and experimentally [29-36]. Among them, as the first successful report for direct epitaxial growth of single crystal STO on Si [1], elemental Sr for the preparation of the buffer layer is important and widely studied [4].

At the cleaved Si (001) surface, each Si atom has *a priori* two dangling bonds pointing out of the surface. Due to the lack of upper bonding partners, pairs of Si atoms prefer to dimerize, forming the well-known "dimer-row" (1 × 2) reconstruction. The addition of half-monolayer (½ ML) Sr buffer layer on this clean Si surface results in a passivated (1 × 2) reconstructed surface alternating rows of Si dimers and Sr atoms. The stability of such an ideal Sr/Si template has been confirmed by first-principles calculations [37-39], as well as several experimental techniques [40-45]. In contrast to this initial Sr/Si (1 × 2) structure with Si dimers, synchrotron x-ray diffraction [24] and scanning transmission electron microscopy measurements after STO deposition [33] indicate a (1 × 1) in-plane structure for the epitaxial STO/Si system with a full-monolayer (1 ML) of Sr at the interface between the STO layer and the Si substrate and no surface Si dimers. It is then natural to ask why, how, and when such a reconstruction of the interface takes place.

In this letter, using first-principles density functional theory (DFT) calculations, we systematically investigate Sr/Si templates and STO/Sr/Si stacks with Sr atoms as an interface buffer layer. The stability and evolution of this buffer layer with STO thickness are analyzed, supporting a reconstruction of the interface during the early stage of the layer-by-layer growth. Properly quantifying the band alignment at the stable interface, we further show that it constitutes a suitable photocathode for water reduction.

*Computational method.—* The DFT calculations are performed using the accurate full-potential projector augmented wave (PAW) method [46], as implemented in the Vienna *ab initio* Simulation Package (VASP) [47-49]. The exchange-correlation potential is treated in the PBEsol [50] form of the generalized gradient approximation (GGA) and a kinetic-energy cutoff of 500 eV is used for the wavefunctions. Regarding the band alignment analysis, we use the Heyd-Scuseria-Ernzerhof (HSE06) hybrid functional [51] to obtain a better description of the electronic structures of bulk Si and STO, and of STO/Sr/Si heterostructures. We consider $12 \times 6 \times 1$ and $18 \times 9 \times 1$ Γ-centered k-point meshes respectively in geometry optimizations and self-consistent calculations. The convergence criterion for the electronic energy is fixed at $10^{-6}$ eV and all structures are relaxed until the Hellmann-Feynman forces on each atom are less than 1 meV/Å. As in previous studies [21,37], the Si substrate is modeled using a five atomic layer Si slab, with the dangling bonds of the Si bottom layer saturated with H atoms to mimic a semi-infinite substrate. An in-plane $(1 \times 2)$ supercell is considered to accommodate the "dimer-row" reconstruction of the Si (001) top surface. A vacuum region of at least 15 Å is included between repeated slabs to eliminate the spurious slab-slab interactions. To prevent spurious electrostatic effects in the treatment of asymmetric structures with periodic boundary conditions, a dipole correction [52] is applied in all simulations.

*The Sr/Si template.*—We start investigating the structural and electronic properties of Sr-buffered Si surfaces. Here we focus on the cases with the deposition of ½ ML (Fig. 1(b)) and 1 ML (Fig. 1(c)) Sr coverage that are relevant for the present study.

As comparison, the cleaved Si (001) surface without coverage is illustrated in Fig. 1(a). At this well-known "dimer-row" $(1 \times 2)$ reconstructed surface, Si atoms possess only three covalent bonds and the surface is thus metallic.

The ½ ML Sr/Si (001) surface, as displayed in Fig. 1(b), is a widely used template layer for growing epitaxial oxides [1]. Due to its $5s^2$ configuration, each Sr atom can provide two electrons to Si. The Sr atoms are located in between the dimer rows and make ionic bonds with Si atoms left and right to saturate their remaining dangling bond. The resulting insulating nature of the surface confirms that the dangling bonds have been successfully passivated by Sr atoms. As already pointed out by Först *et al*. [21], such a Sr-passivated surface with insulating character is therefore a suitable and widely-applied platform for the further growth of STO.

When we adsorb more Sr atoms in order to reach 1 ML Sr coverage, two extra electrons are provided per Sr atom so that we can expect breaking the Si dimers and getting a $(1 \times 1)$ Si surface without "dimer-row" reconstruction. It is indeed what is observed in the top panel of Fig. 1(c). In principle, this 1 ML Sr coverage should passivate the Si surface in a way similar to what was discussed above and yield an insulating template. However, in this case, the charge transfer appears less efficient and the density of states (DOS), as displayed in the bottom panel of Fig. 1(c), clearly shows some metallic characteristics.

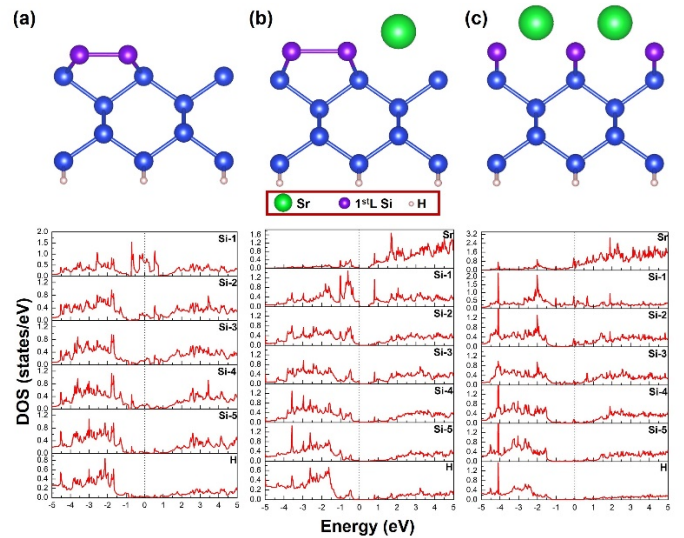

FIG. 1 Influence of Sr coverage on the atomic structure and layer-resolved electronic density of states (DOS) of the Si substrate. (a) Clean dimer-row reconstructed (1 x 2) Si (001) metallic surface without Sr coverage. (b) ½ ML Sr/Si insulating surface preserving (1 x 2) reconstruction. (c) 1 ML Sr/Si metallic surface without dimers. The in-plane lattice constant is constrained as $a$ = 3.843 Å, which gains from the optimized bulk Si. The zero of energy is set as the Fermi level.

*The STO/Sr/Si template.*—We now turn to heterostructures with STO grown on top of the ½ ML Sr/Si and 1 ML Sr/Si systems. The energies differences between the final heterostructures (STO/Sr/Si) and the original templates (Sr/Si) are calculated for the cases with 1 ML Sr ($E_{1ML}$ = -99.450 eV per unit surface, structure S-A) and ½ ML Sr ($E_{½ML}$ = -98.450 eV per unit surface, structure S-B). The interface energy between STO and Si is ~ 1 eV lower in the case with a 1 ML Sr buffer layer, confirming that the $(1 \times 1)$ interface without Si dimers is energetically favored, in excellent agreement with other theoretical and experimental work [24,33].

In particular, experimental measurements reveal that growing STO on a $(1 \times 2)$ Sr/Si template with ½ ML Sr coverage, we end up with a $(1 \times 1)$ STO/Sr/Si stack with 1 ML Sr buffer layer. Two independent scenarios can explain the evolution of the interface. One possibility is that the formation of the 1 ML Sr buffer layer with $(1 \times 1)$ in-plane symmetric Si surface occurs directly when starting the deposition of STO on the ½ ML Sr/Si template. The other possibility is that, as sketched in Fig. 2, an intermediate STO/½ ML Sr/Si state (structure S-B) is formed firstly while Sr atoms diffuse later from the STO bulk or surface to the interface, leaving behind some Sr vacancies.

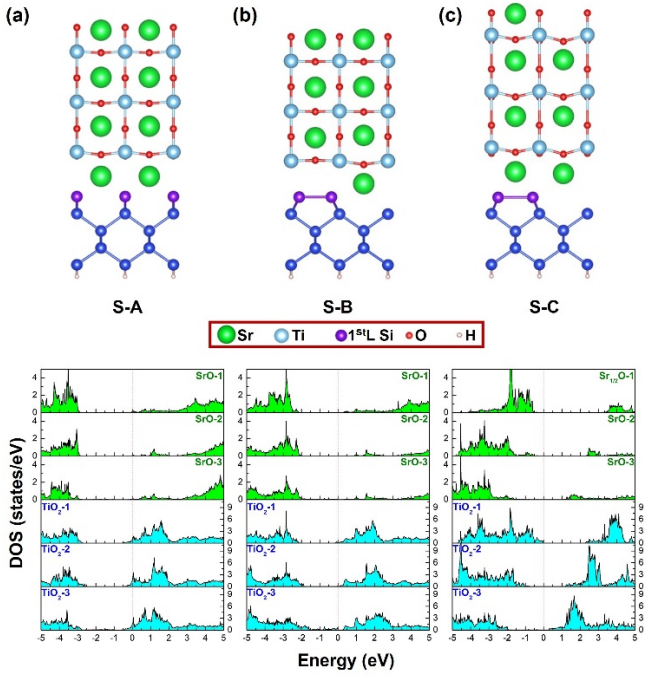

FIG. 2 Sketch of distinct STO/Sr/Si structures labelled (a) structure S-A, (b) structure S-B, and (c) structure S-C and their corresponding layer-resolved DOS for 3ML STO films. The structure S-A – STO/1ML Sr/Si heterostructure in absence of Si dimers is that in agreement of experimental measurements. Directly deposition of STO on ½ ML Sr/Si template would give rise to structure S-B - STO/½ ML Sr/Si heterostructure. The diffusion of Sr from the surface to the interface would then give rise to structure S-C.

In order to probe this second scenario, we construct a heterostructure (structure S-C), in which starting from structure S-B, one Sr atom has moved from the surface to the interface. On the one hand, the relaxed structure S-C shows an energy 0.7 eV per unit surface higher than the original structure S-B and, on the other hand, the interface maintains the (1 × 2) "dimer-row" reconstruction in spite of 1 ML Sr coverage contrary to case S-A. Moreover, compared with the ideal final state S-A, the STO film appears strongly polarized in structure S-C. The averaged rumpling $\delta c_{avg}$ – i.e. the displacement between Sr/Ti and oxygen anions in each atomic layer along the $c$ direction – in structure S-C reaches values up to 0.584 Å, almost 5 times bigger than in structure S-A ($\delta c_{avg}$ = 0.121 Å). Also, the average vertical distance $d_{Sr\text{-}Si}$ between interface Sr and Si planes is computed to be 1.977 Å for structure S-C, larger than 1.607 Å for structure S-A.

The previous observations can be rationalized as follows. STO is a rather ionic material with nominal charges $Sr^{2+}$, $Ti^{4+}$, and $O^{2-}$. In structures S-A and S-B, the $Sr^{2+}O^{2-}$ planes are charges neutral. As clearly displayed in Fig. 2(a) and Fig. 2(b), the DOS of STO film overall maintain its characteristics of the bulk case. In structure S-C, one $Sr^{2+}$ ion has moved from the surface to the interface, so that the surface and the interface acquire respectively a formal charge -2 and +2, giving rise to an electric field pointing from the interface to the surface. The existence of this field is apparent from the inspection of the calculated layer-resolved DOS in Fig. 2(c). It perfectly explains (i) a large electrostatic energy cost for moving Sr from the surface (structure S-B) to the interface (structure S-C) and (ii) the strong rumpling appearing in STO. It also clarifies why the interface Si dimers remain since $Sr^{2+}$ cation cannot provide electrons to the interface (i.e. O in STO is more electronegative than Si).

We further notice that by moving the Sr vacancy progressively from the interface to the surface layer, as shown in Supplementary Material Fig. S1, the rumpling remains consistently confined in the growing region between the buffer layer and the vacancy layer. Also, the energy is progressively increasing consistently with the growing electrostatic energy cost. When the Sr vacancy reaches the surface, there is a partial energy lowering related to the lower formation energy of Sr vacancy at the surface. Still, the lowest energy configuration is that with the Sr vacancy in the buffer layer (structure S-B with ½ ML Sr buffer layer). This proves that $Sr^{2+}$ cations will not move spontaneously from the bulk of the surface to reach 1 ML coverage.

In line with that, if we now artificially dope the system and provide two extra electrons to structure S-C (compensated by a positive background) and fully relax the structure again (structure S-D1), the "dimer-row" reconstruction at the Si surface disappears like in structure S-A. This confirms that dimer opening explicitly relies on the ability of Sr to provide electrons to Si. In Fig. 3(a), the distance between Sr and Si at the interface becomes smaller ($d_{Sr\text{-}Si}$ = 1.701 Å) as well as the rumpling between the plane containing the vacancy and the interface ($\delta c_{avg}$ = 0.407 Å). Inspection of the layer-resolved DOS (the bottom panel of Fig. 3(a)) proves that the build-in electric field has indeed been reduced but is unfortunately not fully suppressed. This is related to the fact that in this artificial calculation the positive background is homogeneously distributed rather than located in a specific layer. Nevertheless, electron doping contributes to compensate the build-in field and is key to open Si dimers.

To model a more realistic situation, we so introduce an O vacancy at the surface of structure S-C, which appears as a concrete and practical way to dope the system and to compensate the build-in field. For the relaxed structure shown as the structure S-D2 in Fig. 3(b) with 1 ML Sr buffer layer, the interface Si dimers successfully open again. Moreover, its layer-resolved DOS (see bottom panel of Fig. 3(b)) reproduces the characteristic of structure S-A in Fig. 2(a), confirming that the build-in electric field is now completely compensated by doping the system through the appearance of oxygen vacancy at the surface. Consistently, the stack exhibits reduced Sr-Si distance ($d_{Sr-Si}$ = 1.612 Å) and rumpling ($\delta c_{avg}$ = 0.149 Å), in good agreement with the values in structure S-A.

It can therefore be questioned if, starting from structure S-B, the formation of oxygen vacancies could eventually favor the migration of Sr from the surface to the buffer layer to reach 1ML Sr interface like in structure S-A. Comparing the energy of structure S-D2 to that of structure S-B with an oxygen vacancy at the surface, it appears however that the latter is almost 0.2 eV per unit surface lower, attesting that Sr will not naturally migrate from the surface to the buffer layer even in presence of oxygen vacancies.

All this supports that when STO is deposited on ½ ML Sr/Si template, the formation of 1 ML Sr buffer layer with (1 × 1) in-plane symmetric Si surface should better happen during the early stage of the layer-by-layer deposition.

It might be additionally questioned if the resulting S-A structure will eventually oxidize giving rise to a SrO interface as for instance suggested in Ref. [14]. This will certainly depend on the experimental conditions and oxygen chemical potential. However, in the case of a SrO interface (as sketched in Supplementary Material Fig. S2), the configuration without dimers (Fig. S2(b), as reported in [14]) is only an unstable stationary point and the stable configuration corresponds to a Si surface keeping dimer rows (Fig. S2(a)), as previously reported by Zhang *et al*. [31] and Chrysler *et al*. [53]. This stable structure is expected from the previous discussions since the ability to open Si dimers relies on the capacity of Sr to provide electrons to Si. In this case, Sr interfacial atoms will preferably provide electrons to O, much more electronegative, rather than to Si.

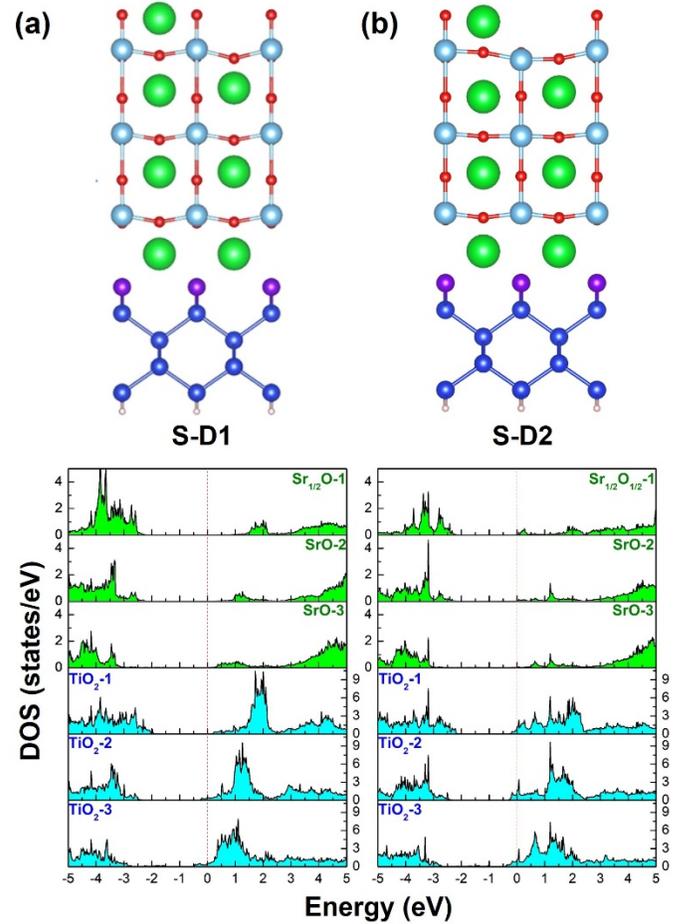

FIG. 3 (a) Starting from structure S-C, the addition of two electrons gives rise to the fully relaxed structure S-D1 without Si dimers. (b) Starting from structure S-C, the addition of an oxygen vacancy at the surface gives rise to the fully relaxed structure S-D2 owing characteristics as in structure S-A.

We so conclude that the STO/Sr/Si interface as currently observed with 1ML Sr coverage on and no Si dimers is our non-oxidized S-A structure.

*Electronic and photocatalysis properties.*—Having established that structure S-A is the one experimentally observed, we now turn to the characterization of its electronic properties. In Fig. 4(a) we report the band alignment obtained with the PBEsol functional. A so-called staggered gap configuration (type II) with valence band maximum (VBM) of Si in the gap of STO and the conduction band minimum (CBM) of STO in the gap of Si. As usual in GGA, both bulk bandgaps are however significantly underestimated. Correcting this in first approximation by applying a naïve rigid shift of the conduction bands ("scissors correction") as frequently done in the literature [21,26,31], the situation evolves toward a straddling gap situation (type I) with valence and conduction band edges of Si now located within the bandgap of STO (dashed lines in Fig. 4(a)).

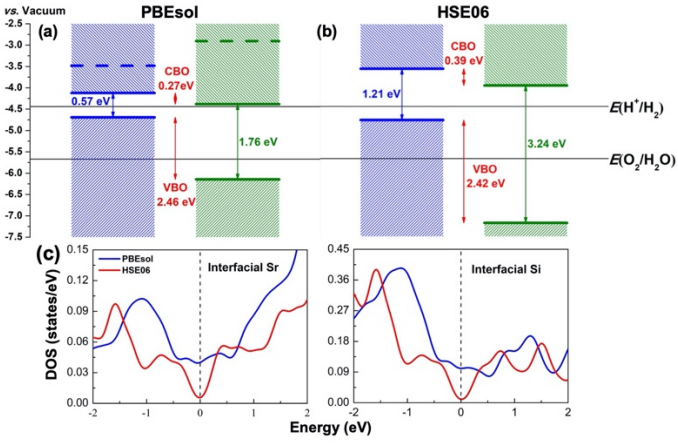

FIG. 4 Band alignment diagrams of the structure S-A using (a) PBEsol and (b) HSE06 functional. The dashed lines in (a) are the estimated CBM positions after an empirical rigid shift to reproduce experimental bandgaps of bulk Si and STO. (c) DOS around the Fermi level taken as energy reference for the interfacial 1ML Sr and Si layers of the structure S-A using PBEsol (blue lines) and HSE06 (red lines) functional.

In order to provide a more trustable description, we perform additional calculations using the hybrid HSE functional that properly reproduces the bandgaps of bulk STO and Si (~ 3.25 eV for STO [54] and 1.12 eV for Si). The results are reported in Fig. 4(b) and confirm a type II configuration with a valence band offset VBO = 2.42 eV and a conduction band offset CBO = 0.39 eV. This reveals that, on the one hand, the band alignment obtained in GGA is qualitatively correct (type II) while, on the other hand, the rigid shift of the conduction bands, as currently done in previous work, should better be avoided since most of the correction concerns the re-alignment of the valence bands.

Assessing the correct band alignment is important since it clarifies that photoexcited electrons and holes can be efficiently separated at this interface. From Fig. 4(b), photogenerated electrons (holes) in Si (STO) can naturally transfer to STO (Si), which will result in the accumulation of electrons in the CB of STO and holes in the VB of Si. Furthermore, aligning CBM and VBM as well as the redox potentials of $H^+/H_2$ and $O_2/H_2O$ on the vacuum, the potential of $H^+/H_2$ appears below the CBM position of STO, indicating that the interface is suitable for water reduction. De facto, the interface appears inappropriate for oxidation in view of the small bandgap of Si.

All this highlights the concrete interest of the stable S-A interface as a photocathode for water reduction. However, the DOS (Fig. 4(c)) points out the existence of interface states, providing a slightly metallic character to the interface, perfectly in line with what is observed at the 1 ML Sr/Si surface, as shown in Fig. 1(c). The description is again qualitatively similar in PBEsol and HSE functionals, both of which reproduce a similar small DOS at the Fermi energy. This finite DOS is however significantly reduced in the more accurate hybrid functional calculations. The Fermi energy is located in a pseudogap region and the finite DOS is related to a few interface states of Sr-$d_{xy}$ character on the Sr buffer layer and Si-$p_y$ character on the Si side.

It might be questioned if this could eventually hamper photocatalytic applications. Experimentally, Ji *et al.* [14] successfully realized already water splitting at a STO/Sr/Si(001) system. In their work, although starting from a ½ ML Sr/Si template, they suggest (in Fig. 2 of Ref. [14]) a fully oxidized interface with 1 ML SrO and no dimers, as in Supplementary Material Fig. S2(b). However, they do not support the presence of oxygen in the buffer layer and, from our previous discussion, this oxidized interface without dimers is unstable. Moreover, this interface with negligible CBO would be insufficient for electron accumulation. So, it is very likely that their interface corresponds to our stable structure S-A. In agreement with that, they report the presence of interface states but conclude that their density is small enough to avoid preventing water splitting applications.

We further notice that, as illustrated in Fig. S2(a) and Fig. S2(c), changing experimental conditions to favor a fully oxidized interface would give rise to a truly insulating structure with dimers but showing however a type I band alignment inappropriate for charge separation. So, although the S-A structure as conventionally obtained without dimers shows a small density of interface states, it remains the most promising for water splitting applications.

*Conclusions.*—In summary, using first-principles density functional theory calculations, the interface of STO with Si in presence of a Sr buffer layer has been systematically investigated. Regarding first the initial Sr/Si template, we confirm that the ½ ML Sr/Si structure with a "dimer-row" (1 × 2) reconstructed Si surface is an ideal insulating platform for the epitaxial growth of STO film. The 1ML Sr/Si structure shows a stable (1 × 1) Si surface without dimers but exhibits a weak metallic character. Regarding then the final STO/Sr/Si stack, the interface between Si and STO with 1 ML Sr buffer layer is energetically favored, in line with experimental findings. This STO/1ML Sr/Si structure has no Si dimers and exhibits a small density of interface states at the Fermi level. Starting from a STO/Sr/Si stack with 1/2 ML Sr buffer layer, we see no natural energetic reason for Sr to migrate to the interface. Migration of $Sr^{2+}$ would give rise to an electrostatic energy cost and would not be compatible with the disappearance of Si dimers. Moreover, even in presence of oxygen vacancies that help to suppress the electrostatic energy cost and open Si dimers, the Sr migration to the buffer layer remains energy costly. This supports the scenario that the reconstruction of Sr buffer layer from the initial ½ ML configuration to the final 1 ML configuration experimentally observed in the STO/Sr/Si stack takes place during the early stage of the growth process. We argue also that the interface experimentally observed should be non-oxidized. Further characterizing the electronic properties of this interface, we point out its type II band alignment appropriate for charge separation and confirm the potential interest of this interface as photocathode for water reduction. We hope this work will motivate further experimental exploitation of this interface.

W.T. and E.B. acknowledge individual supports from F.R.S.-FNRS Belgium. M.S. acknowledges funding from Slovenian Research Agency (Grants No. N2-0176, N2-0187 and P2-0091). This work was also supported by the SIOX project from ERA.NET and PROMOSPAN PDR grant from F.R.S.-FNRS Belgium. The authors acknowledge access to the CECI supercomputer facilities funded by the F.R.S-FNRS (Grant No. 2.5020.1) and to the Tier-1 supercomputer of the Federation Wallonie-Bruxelles funded by the Walloon Region (Grant No. 1117545).

*Corresponding author.
Philippe.Ghosez@uliege.be

# Supplementary Material - First-principles investigation of interfacial reconstruction in epitaxial SrTiO$_3$/Si photocathodes

Wen-Yi Tong[1], Eric Bousquet[1], Matjaž Spreitzer[2] and Philippe Ghosez[1,*]

[1] *Theoretical Materials Physics, Q-MAT, CESAM, Université de Liège, Sart Tilman B-4000, Belgium*
[2] *Advanced Materials Department, Jozef Stefan Institute, Jamova 39, 1000 Ljubljana, Slovenia*

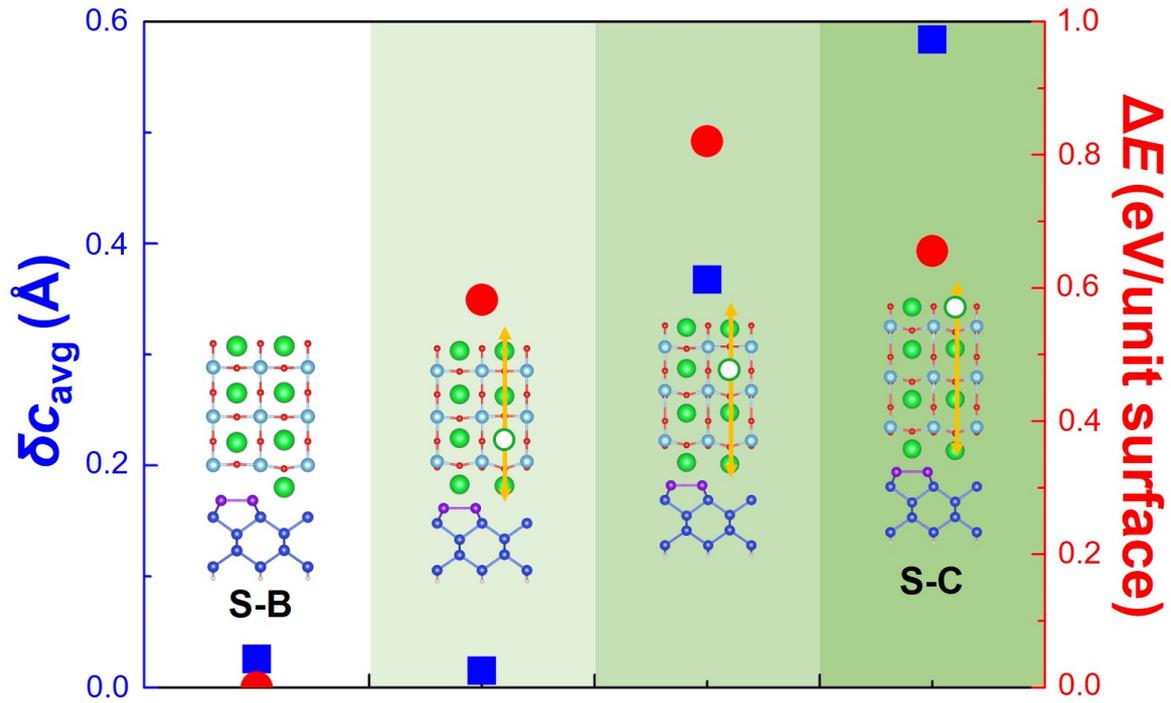

**FIG. S1 - Evolution of the rumpling and energy as a function of Sr vacancy position.** During the motion of Sr from the interface to the surface, the polarization of the STO film $\delta c_{\text{avg}}$ (blue squares), and the total energy (red circles) respect to the structure S-B taken as reference are strongly evolving as a consequence of the build-in electric field (yellow arrows) related to Sr vacancies.

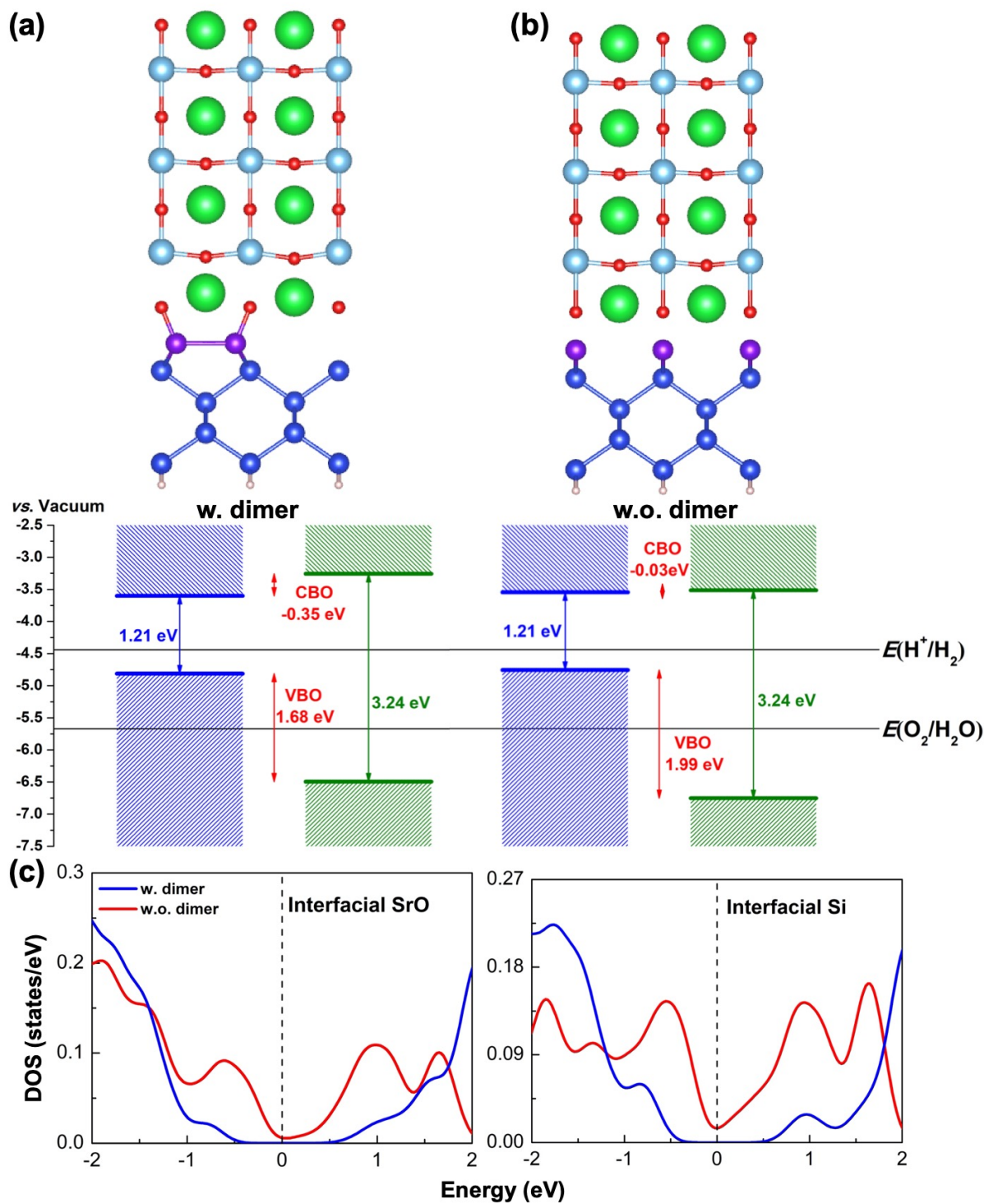

**FIG. S2 - Sketch of STO/SrO/Si structures and their band alignment diagrams** (a) with dimers and (b) without dimers. The structure (a) is 0.544 eV per unit surface lower in energy than the structure (b). The band edge positions (blue for Si and green for STO) with respect to the vacuum level are obtained through core-level alignment scheme. The vacuum level is determined through calculating the averaged electrostatic potential along the $z$ axis in the slab model. The Fermi level $E_f$, redox potential of normal hydrogen electrode $E(H^+/H_2)$ (~ -4.44 eV at pH = 0) and the oxidation potential $E(O_2/H_2O)$ (~ -5.67 eV at pH = 0) are labelled. (c) Density of states around the Fermi level taken as energy reference for the interfacial SrO and Si layers of the structure with dimers (blue lines) and without dimers (red lines).